\documentstyle[aps]{revtex}
\input epsf.tex
\def \Laa{$L_{a}$}

\def \beq{\begin{equation}}
\def \eeq{\end{equation}}
\begin{document}
\baselineskip=24pt
\begin{center}
\bf{Finite Temperature Properties of the SO(3) Lattice Gauge Theory } \\

\vspace{1cm}

\rm{Srinath Cheluvaraja}\footnote[1]{e-mail:srinath@theory.tifr.res.in} \\
Tata Institute of Fundamental Research \\
Mumbai 400 005,
India
 and \\
\rm{H.S. Sharathchandra}\footnote[2]{e-mail:sharat@imsc.ernet.in} \\
The Institute of Mathematical Sciences \\
Madras - 600 113, INDIA\\
\end{center}
\vspace{1cm}

\noindent{\bf{ABSTRACT}}\\
We make a numerical study of the finite temperature properties of the 
$SO(3)$ lattice gauge theory. 
As its symmetry properties
are quite different from those of the $SU(2)$ LGT, a different
set of observables have to be considered in this model.
We study several observables, such as, the plaquette square, the $Z(2)$ monopole
density, the fundamental and adjoint Wilson line, and the tiled Wilson
line correlation function. Our simulations show that
the $Z(2)$ monopoles condense at strong coupling
just as in the bulk system. This transition is seen at approximately
the same location as in the bulk system.
A surprising observation is the multiple valuedness
of the adjoint Wilson line at high
temperatures. At high temperatures, we observe long lived metastable states in which
the adjoint Wilson line takes positive and negative values.
The numerical values of other
observables in these two states appear to be almost the same.
We study these states using different methods and also make comparisons
with the high temperature behaviour of the $SU(2)$ LGT.
Finally, we discuss various interpretations of our results and point out
their relevance for the phase
diagram of the $SO(3)$ LGT at finite temperature.

\vspace{0.5cm}
\begin{flushleft}
\end{flushleft}

\newpage
\begin{section}{Introduction}
Lattice Gauge Theories (LGTs) at non-zero temperatures have been studied
extensively for many years. 
They have provided us with models for the confinement-deconfinement phase transition
which is expected to occur in realistic theories like Quantum
Chromodynamics.
The 
thermodynamic properties of the $SU(2)$ and
$SU(3)$ LGTs have also been vigorously studied \cite{thmics}.
Nevertheless, the implications of LGTs
for continuum Yang-Mills theories are not completely
clear. There are many questions
about the high temperature phase
which have still eluded an understanding; some of these are:
a precise characterization
of the phase, the structure of its elementary
excitations, and its static and dynamic properties.
This makes the study of the finite temperature properties of LGTs a subject
of continuing interest.

The pioneering work in \cite{suss} was the first non-perturbative
calculation to show that quarks are deconfined at high temperatures.
The analysis in \cite{suss} is done in the strong coupling limit 
( $g>>1$) 
of the $SU(2)$ LGT. In this limit,
the partition function of the $SU(2)$ LGT 
is rewritten as a 3-d spin model
with a global $Z(2)$ symmetry. The ordered phase of this spin model
corresponds to the deconfined phase 
and the disordered phase corresponds
to the confined phase.
Following this calculation, Monte-Carlo simulations \cite{early}
provided further evidence that the transition 
takes place in the physical
weak coupling limit ($g <<1$).
These simulations are usually done (for the $SU(2)$ LGT) 
using the Wilson action \cite{wils} which is defined as
\beq
S=\frac{\beta_{f}}{2}\sum_{n\ \mu < \nu}tr_{f}U(n\ \mu \nu) \quad .
\eeq
The $U(n\ \mu \nu )$s are the plaquette variables that are
the oriented product of the $SU(2)$ link variables along an elementary
square and are constructed as:
\beq
U(n\ \mu \nu)=U(n\ \mu)U(n+\mu\ \nu)U^{\dagger}(n+\nu\ \mu)
U^{\dagger}(n\ \nu) \quad .
\eeq
$U(n\ \mu)$ are the $SU(2)$ variables defined on the links.
The basic observable that is studied
in simulations is the Wilson-Polyakov line (henceforth called the Wilson line)
; this is defined as
\beq
L_{f}(x)=Tr_{f}P\ \exp i\int_{0}^{\beta}A(x)dx_{4} \quad .
\eeq
The subscript $f$ indicates that the trace is taken in the
fundamental representation of the group.
In analogy with the Wilson loop,
the expectation value of this observable can be interpreted as the free energy of a
static quark in a heat bath (at a temperature $\beta^{-1}$).
This connection is made explicit by writing it in the form:
\beq
<L_{f}(x)>=\exp (-\beta F(x)) \quad .
\eeq
A non-zero expectation value of the Wilson line signals deconfinement; a zero
expectation value signals confinement. This observable is the order parameter
for studying the confinement-deconfinement phase transition in $SU(N)$ LGTs.
The importance of the center of the gauge group,$Z(N)$ for $SU(N)$, was further
underlined in \cite{yaffe} where it was proposed that the critical behaviour of
4-d $SU(N)$ gauge theories could be understood in terms of the critical behaviour
of 3-d spin models having a global $Z(N)$ symmetry. The group $Z(N)$, which
is the center of the group $SU(N)$, plays a special role in the
deconfinement transition. This is because of an extra symmetry in the
finite temperature gauge theory that arises from the periodic
boundary conditions in the Euclidean time direction; gauge transformations
which are periodic (in time) upto a constant center element leave the
action invariant.
The Wilson line picks up a phase under the action of these gauge
transformations; it transforms as
\beq
L_{f}(x) \rightarrow Z L_{f}(x) \quad ;
\eeq
here Z is an element of the center, and
for the group SU(2), it is either $+1$ or $-1$.
Therefore, a non-zero value of the Wilson line at high temperatures 
signals a spontaneous breaking
of the global center symmetry,
implying
that the high temperature phase is degenerate,
with the two degenerate states related by a Z transformation.
Numerical simulations of the $SU(2)$ LGT observe
these degenerate states as 
metastable states in simulations. At high temperatures,
the Wilson line  settles to either a
positive or a negative value and remains in either of these two states for
very long simulation times.
The order of the transition to the high temperature phase
has also been investigated thoroughly in the $SU(2)$ and $SU(3)$ LGTs.
The expectations in 
\cite{yaffe}, concerning the order of the phase transition, 
have been borne out for $SU(2)$ \cite{su2is}
and $SU(3)$ \cite{mont} LGTs in which one observes a 3-d Ising like critical
behaviour and a 3-d Z(3) like first order transition, respectively.

Since lattice actions are anyway not unique, it is natural to study the finite
temperature properties of LGTs using equivalent actions. 
The universality of lattice gauge theory actions requires that different
actions, which correspond to different regularizations of Quantum Field
Theories,
 should reproduce the same physics in the continuum limit.
One such LGT is
defined by
\beq
S=\frac{\beta_{a}}{3}\sum_{p}tr_{a}U(p) \quad .
\label{so3}
\eeq
Unlike the Wilson action, the trace of the plaquette is taken in the adjoint
representation. The trace in the two representations are related by
\beq
tr_{a}\ U=(tr_{f}\ U)^2 -1 \quad .
\eeq
Though this action is defined using $SU(2)$ link variables
and the $SU(2)$ Haar measure, it describes an $SO(3)$ LGT because
the link variables $U(n,\mu)$ and $-U(n,\mu)$ have the same weight.
In this paper, we will report on our studies with this action and we will
encounter some unexpected and interesting phenomena.

There are several reasons why a study of the finite
temperature properties of the $SO(3)$ LGT can be useful and important.
The $SO(3)$ LGT
has the same naive continuum limit as the $SU(2)$ LGT, and is
expected to lead to the same physics as the $SU(2)$ LGT.
Furthermore, since the group $SO(3)$
has no non-trivial center subgroup like $SU(2)$, it would be interesting to see
how it can reproduce the same properties as the $SU(2)$ LGT in the
absence of a non-trivial center subgroup.
Also,
unlike the $SU(2)$ LGT, the $SO(3)$ LGT has a first order bulk transition
at $\beta_{a}\approx 2.6$ that is driven by the condensation of $Z(2)$ monopoles
\cite{hall}. The condensation of these $Z(2)$ monopoles has nothing to do with
deconfinement. Both sides of the bulk transition are confining phases, and only the
$Z(2)$ degrees of freedom behave differently in these two phases. The
presence of these additional $Z(2)$ degrees of freedom should lead to a
richer phase diagram in which both sides undergo phase transitions into
a high temperature phase.
Another issue which has been discussed recently is the difficulty in separating a
bulk and a finite temperature transition. The finite
temperature properties of the mixed action $SU(2)$ LGT, that was 
defined in \cite{creu},
were recently studied in \cite{gav} and it was found that
the deconfinement transition joined the bulk
transition making it difficult to separate the two. This
raises the issue whether it is possible to make any meaningful distinction
between these two transitions. Similar studies have also been made with a mixed
action $SU(3)$ LGT in \cite{blum}.
It is with these motivations in mind that we have tried to understand the finite
temperature properties of the $SO(3)$ LGT.
The Monte-Carlo simulation method is used to arrive at the numerical
results.
We run into many puzzling features in our studies of the $SO(3)$ LGT.
Our studies indicate that the $SO(3)$ LGT has a much richer behaviour than
the $SU(2)$ LGT.
In this paper we first present our numerical observations and
later make proposals for their physical interpretation. 
We consider different scenarios for the phase diagram of the $SO(3)$ LGT.
Though we do not have a convincing proof for any particular scenario, we
present several reasons for favouring the scenario we believe is true.

We first observe that the Wilson line in the fundamental representation
is not an order parameter in the $SO(3)$ LGT. This is because the
global $Z(2)$ symmetry present in the $SU(2)$ LGT is promoted to a local
symmetry in the $SO(3)$ LGT. The center transformation can now depend on
the spatial position and acts as:
\beq
\L_{f}(x)\rightarrow Z(x) \L_{f}(x) .
\label{local}
\eeq
Since local symmetries are never spontaneously broken \cite{elitz},
this forces the average value of the  
Wilson line in the fundamental representation
to be always zero. Only observables which are invariant under this local
symmetry can have a non-zero average value in this model. Before we discuss
these observables, it is illuminating to rewrite the action for the
$SO(3)$ LGT in a slightly different form. This involves linearizing
the square term of the trace by
introducing an auxillary gaussian field ($\lambda(p)$) on the plaquettes,
after which the action becomes
\beq
S=\frac{\sqrt \beta_{a}}{\sqrt 3}\sum_{p}\ tr_{f}U(p)\lambda(p)- 
\frac{1}{4}\sum_{p}\lambda(p)^2 \quad .
\label{niceform}
\eeq
In the above form, the $SO(3)$ LGT is like an $SU(2)$ LGT interacting with
additional gaussian plaquette degrees of freedom. These plaquette variables are
the $Z(2)$ degrees of freedom. This form
also shows that the $SO(3)$ LGT has additional degrees of freedom 
compared to the
$SU(2)$ LGT. The $SU(2)$ LGT is recovered when the additional $Z(2)$
variables are frozen to +1.
The above form of the action, 
unlike the form in Eq.~\ref{so3},
is also convenient for simulations for which a heat bath or an
overrelaxation algorithm can be easily implemented. 
The action has the local $Z(2)$ invariance:
\beq
U(n,\mu)\rightarrow -U(n,\mu)\ \  
\lambda(p)\rightarrow -\lambda(p) \quad;
\label{loc}
\eeq
the $\lambda(p)$s being the plaquettes touching the link $U(n,\mu)$.
To study this model, we must construct observables which are invariant
under these local gauge transformations. Wilson loops and Wilson-Polyakov
lines fail to satisfy this criterion and their average values 
are identically zero.
Nevertheless, we can discuss the behaviour of several observables which are
invariant under these local gauge transformations.
One such observable is a sheet variable.
An example of a sheet variable is
the "tiled" Wilson loop
\beq
W(C)=\prod_{l\in C}U(l) \prod_{p\in C}\lambda(p) .
\eeq
The first part of the observable is the usual Wilson loop 
defined over a loop $C$, and the
other part consists of the auxillary $Z(2)$ variables which are defined
on all the plaquettes enclosed by the loop $C$.
The tiled Wilson loop cannot be given the usual physical interpretation of
the potential of a 
quark-antiquark pair because additional $Z(2)$ degrees of
freedom are involved in its definition. Nevertheless, it is an interesting
gauge invariant variable which incorporates both the $SU(2)$ and the $Z(2)$
degrees of freedom in the $SO(3)$ LGT.
Similarly, we can define a  "tiled" Wilson line correlation function  as
\beq
W(x,y)=tr\ L_{f}(x)L_{f}(y) \prod_{p\in C}\lambda(p) .
\eeq
We expect this observable to be useful in studying the finite temperature
properties of the $SO(3)$ LGT.
The $Z(2)$ monopole density, $\rho$, can be extracted from the 
$\lambda(p)$ variables as follows:
\beq
\rho(c)=\frac{1}{2}(1-sgn(\prod_{p\ p\in \partial c}\lambda(p))) \quad .
\eeq
This definition of the monopole density is also gauge invariant.
From its definition,
a $Z(2)$ monopole is present in a 3-d cube
whenever the product of the $Z(2)$ auxillary variables bordering the
cube is negative.
The $Z(2)$ monopoles can be imagined as lattice monopole configurations
that carry
a net $Z(2)$ magnetic flux. 
Another observable of interest is the
Wilson line in the adjoint representation that is defined as
\beq
L_{a}(x)=Tr_{a}P\ \exp i\int_{0}^{\beta}A(x)dx_{4} \quad .
\eeq
This observable can also be studied in the $SU(2)$ \cite{adjoint,kiss,thesis}
and the $SU(3)$ \cite{faber} LGTs at finite temperature, and it can be used
to monitor the deconfinement transition. However, 
the adjoint Wilson line
plays a much more essential role in the study of deconfinement in the
$SO(3)$ LGT.
For the group $SU(2)$,
this observable can be expressed in terms of
the fundamental Wilson line $L_{f}$ by the relation
\beq
L_{a}=L_{f}^2-1 \quad  .
\eeq
In this form, it is easy to see that the adjoint Wilson line is
invariant under the local $Z(2)$ transformation in Eq.~\ref{local} and
is in general non-zero.
In analogy with the fundamental Wilson line in the $SU(2)$ LGT, we expect
the adjoint Wilson line to tell us something about the deconfinement
transition in the $SO(3)$ LGT. The physical interpretation attached to the
fundamental Wilson line carries over to the adjoint Wilson line. It measures
the free energy of a static source ($F_{a}$)
in the adjoint representation placed inside a heat bath at a temperature $\beta^{-1}$.
This is again seen by writing it as
\beq
\label{freen}
\langle L_{a}(x) \rangle =\exp (-\beta F_{a}(x)) \quad .
\eeq
A non-zero value of this observable implies that such a static source
has a finite free energy.
It must be noted, however, that confinement of adjoint sources is to be understood
slightly differently from confinement of fundamental sources. An adjoint source
(which is a non-abelian charge in the $j=1$ representation of $SU(2)$)
can always bind with two fundamental sources ($j=1/2$) and form a
colour singlet bound state. Similarly,
two widely separated adjoint sources will form
two colour singlet bound states without any string joining the two.
Hence, unlike the fundamental Wilson line, the adjoint Wilson line is always
non-zero, and it is not an order parameter in the strict sense.
Nevertheless, it can show discontinuous behaviour across a phase transition
just like any other observable.
Since the behaviour of an adjoint source depends on its ability to
bind to fundamental sources, we expect the adjoint source to closely follow
the behaviour of fundamental sources. This is true for the
$SU(2)$ LGT in which the adjoint Wilson line can 
equally well be used to
locate the deconfinement transition.
Finally, the other gauge invariant variable we consider is the square
of the plaquette variable defined as
\beq
P= (1/3)tr\ U(p)^2 \quad .
\eeq
This measures the energy density in a bulk sysytem.

In the next section we present our numerical studies
of the above mentioned observables, and then we attempt to
provide a physical interpretation to our results.
\end{section}
\begin{section}{Numerical Results.}
In this section we present our numerical results. We first briefly
describe the systematics of the simulation.
A metropolis update (with 3 hits)  followed by overrelaxation updates (2 hits)
was
used to generate the configurations. Measurements were made every
ten sweeps after omitting the first thousand configurations.
We performed runs ranging from 10000 to 50000 Monte-Carlo
sweeps.
The link variables and the
gaussian variables were updated separately. 
Since any simulation should also incorporate the local invariance
in Eq.~\ref{loc}, the transformations in Eq.~\ref{loc} were implemented
every time a measurement was made.
The link variables were updated
by multiplying them with an $SU(2)$ element chosen at random from a table
consisting of 50 elements which was biased to lie close to the unit element.
The auxillary gaussian variables were updated by adding to them a number
randomly chosen in the interval $(-cut,cut)$. The value of cut was chosen 
so that an acceptance rate of around
forty percent was roughly maintained for both the updated variables.
A finite temperature simulation is mimicked by choosing a lattice of
small temporal extent, a large spatial extent, and periodic
boundary conditions ( with period $\beta^-1$) in the temporal direction. We have
made our studies on lattices of different sizes.
The maximum spatial size used was $N_{\sigma}=10$ and
the maximum temporal size was $N_{\tau}=7$. Unless otherwise mentioned,
the lattice size is usually $7^3\ 3$.

We have decided to present our numerical results first along with
some explanations, and only in the end do we start giving our
interpretations. Though this may appear a bit tedious, 
there are some reasons for doing this.
The numerical
results are interesting in their own right and many of them are quite
unexpected. Even before we discuss matters of interpretation, the
numerical observations themselves present some puzzling features.
The other reason for this approach is that the numerical results can
always be considered separately from any physical interpretation we
wish to attach to them; they can be regarded as empirical
observations that have to be properly explained.

The
observables that were studied were the plaquette square, the
$Z(2)$ monopole density, the adjoint and fundamental Wilson line, 
the tiled Wilson line correlation function, and the auxillary
$Z(2)$ variable. We shall discuss these in turn. 
We shall use the terms small $\beta_{a}$
and large $\beta_{a}$ interchangeably with low temperature and high temperature,
respectively.

The $Z(2)$ monopole density and the plaquette square show an abrupt
change at $\beta_{a}\approx 2.5$. 
Fig.~\ref{trans} shows the behaviour of these two observables.
The discontinuous jump in these quantities suggests a first order
transition just as is observed in the bulk system.
There is no indication of any other phase
transition.
The abrupt change in these observables
signals
a phase transition between the two regimes $\beta_{a}<2.5$ and
$\beta_{a}>2.5$. The region $\beta_{a}<2.5$ is a condensate of
$Z(2)$ monopoles while the region $\beta_{a}>2.5$ has virtually
zero monopole density. Even the location of the phase transition,
$\beta_{a}^{cr}\approx 2.5$, is
very close to that observed in the bulk system  ( $\beta_{a}^{cr} \approx 2.6$).

We now come to the behaviour of the
Wilson line in the adjoint representation ($L_{a}$) which is the most
interesting aspect of our studies.
At low temperatures, it remains very small, but it jumps to a
non-zero value at high temperatures. This jump occurs across
$\beta \approx 2.5$ which is the same point
where the plaquette square and the $Z(2)$
monopole density show a discontinuous behaviour.
The startling feature is that the
adjoint Wilson line takes two distinct values at high temperatures.
Depending on the starting configuration of the Monte-Carlo run, the
adjoint Wilson line takes either a positive or a negative value.
A cold start ( corresponding to an initial configuration where all the link
variables are unity ) always leads to the state with \Laa \ positive, while
a hot start (corresponding to a initial configuration where all the link
variables are randomly distributed) usually leads to the state with \Laa \ 
negative.
The two metastable states are shown in Fig.~\ref{la}(a).
The reason why we call them metastable states will be explained later.
In order to test whether these states are truly long lived
metastable states the updating algorithm was tampered with in various ways,
but these states always appeared. Infact, the raison d' etre 
for simulating the
action in Eq.~\ref{niceform} was to design an overrelaxation
algorithm which could be used to verify these metastable states.
Another surprising feature is that the average value of the plaquette
square observable
in both these metastable states appears to be 
almost (but not exactly) equal. The value of the $Z(2)$
monopole density is exteremely small at high temperatures and is not
significantly different in these two metastable states.
We also mention that we have hardly been able to see any tunnellings between
these two metastable states except in a situation to be described later.
A plot of \Laa \  vs $\beta_{a}$ is presented in Fig.~\ref{la}b. At high
temperatures, we show the values of \Laa \ in both the metastable states 
which are observed
in simulations.

Before accounting for these metastable states, 
we take a look at the distribution functions (normalized to one)
of the adjoint and fundamental Wilson line at high and low temperatures.
They will help us to understand the structure of the high and low
temperature states.
We have plotted the single site distribution function because that
gives more information about the configurations in these states.
Fig.~\ref{plus} and Fig.~\ref{minus} show these distributions for the
the two high temperature states.
In the state with \Laa \  negative, there is a sharp peak at
$-1$ while the state with \Laa \  positive is peaked at a positive
value (close to +3).
We have already noted that the fundamental Wilson line will always have
a zero expectation value because of the local $Z(2)$ symmetry. This
requires
\beq
\langle L_{f}(x) \rangle=0 \quad .
\eeq
A zero expectation value can arise in different ways. Either the values
of $L_{f}$ can be peaked about zero or there can be two peaks at
non-zero values symmetrically distributed about zero. The distribution
of $L_{f}$ in the two high temperature states shows that both these
possibilities occur. The \Laa \ positive state has double
peaks symmetrically placed about zero; the \Laa \ negative state
has a sharp peak about zero.
At low temperatures, the distribution of
$L_{f}$ is broadly peaked about zero. The distribution of $L_{a}$
shows a peak at $-1$ but there is a tail stretching all the way
to $3$.
These are shown in
Fig.~\ref{zero}.
In Fig.~\ref{su2high} 
we also display
similar distributions for the
$SU(2)$ LGT at high temperatures.
The distribution of $L_{a}$ in the 
\Laa \ positive
state is very similar to its distribution 
in the high temperature phase of the
$SU(2)$ LGT.
At low temperatures, the 
distributions of $L_{f}$ and $L_{a}$ in the
$SU(2)$ theory are very similar to those in the $SO(3)$ theory and so we 
donot present them.
From these plots, the state with \Laa \ 
positive in the $SO(3)$ LGT is seen to be quite similar to the high temperature 
(deconfined) phase of the
$SU(2)$ LGT. 
The state with \Laa \ negative is, of course, absent in the $SU(2)$ 
LGT.
Let us now compare the distribution of the adjoint Wilson line in the
\Laa \ negative state with the low temperature, \Laa \ 
$\approx 0$,
state.
Both the
profiles are peaked at $L_{a}\approx -1$, but there is a tail extending
all the way to +3 in one, whereas in the other, the tail is truncated very
sharply. A similar comparison of the $L_{f}$ distribution shows that
the two states differ only by the sharpness of their peaks centered
on $0$.
From the above observations, we conclude that although we see only one
phase transition at $\beta_{a}\approx 2.5$, and this transition involves
only the $Z(2)$ degrees of freedom, the distribution functions of
the fundamental and adjoint Wilson line are sufficiently modified
across the transition. The \Laa \ negative state at high temperature
and the low temperature state are quite similar insofar as their
configurations, which are peaked about $L_{a}=-1$; only the width 
of the distributions are
different in the two cases. 
On the other hand,
the \Laa \ positive state 
has a peak at a different location. Likewise, the distribution of
$L_{f}$ in the low temperature phase differs from the one in the 
\Laa \ 
negative state only by the width of its peak.
In the \Laa \ positive
state, however, the $L_{f}$ distribution is quite different and has two double
peaks symmetrically placed about zero.

Another observable which can also be
monitored is the average of the auxillary plaquette variable. This observable is
not
gauge invariant and represents the additional $Z(2)$ degrees of freedom
in the $SO(3)$ LGT. It is similar to $L_{f}$ because it 
can arbitrarily flip its sign giving it a zero average value. 
Though its average value is always zero, its distribution undergoes
a change across the transition just as $L_{f}$. This is 
shown in Fig.~\ref{sigdis}.
At low temperatures, it has a broad peak centered on zero; at high
temperatures, it has two peaks placed symmetrically about the origin.
This means that the higher moments (like $\lambda(p)^2$ which is gauge
invariant)
will show a discontinuous behaviour
across the transition.

Now we turn to another aspect of our numerical results.
We had mentioned earlier that in the states with \Laa \ negative 
and \Laa \ 
positive, the values of observables like the plaquette square and
the $Z(2)$ monopole density were almost equal. This should be checked
for different values of $\beta_{a}$ and $N_{\tau}$. On an $N_{\tau}=3$
lattice, the two states have the same value of the plaquette square 
observable for
a wide range of couplings. The differences
between the two states start showing up only at very large couplings.
We plot the time evolution 
(Fig.~\ref{evolu}) of the plaquette square observable 
for these two metastable states on
an $N_{\tau}=3$ lattice for two different couplings, $8.5\ and\ 10.5$.
We notice that the two values start moving apart only after $\beta_{a}=
8.5$. The same feature is observed when we go to lattices of temporal
size $N_{\tau}=2$. This means that at very high temperatures (large $\beta_{a}$
or small $N_{\tau}$), the \Laa \ positive and 
\Laa \ negative states
begin to differ slightly from each other at least in the values of
the plaquette square observable ( which is the energy density
in a bulk system).

The tiled Wilson line correlation function also behaves differently at low
and high temperatures. At low temperatures, it falls rapidly to zero at 
large distances; in the two high temperature phases, it again
behaves differently; in the \Laa \ positive state, it reaches a
non-zero value at large distances, and in the \Laa \  negative state,
it falls to zero at large distances
just as in the low temperature phase.
This is shown in Fig.~\ref{tiled}. 
This measurement of the correlation function was done on a $10^3\ 3$ lattice.
Some simple arguments can be given for this behaviour of the tiled
Wilson line correlation function.
At strong coupling, it is natural to expect the tiled
Wilson line correlation function to have an area law. This translates into a rapid fall
of the correlation function.  At high temperatures, a
different argument can be made. The very small (virtually zero)
$Z(2)$ monopole density means that virtually all the cubes in the
lattice satisfy
\beq
\sigma(c)=+1 \quad.
\eeq
This condition can be satisfied by
\beq
\lambda(p)=|\lambda| \prod_{l\in \partial p} z(l) \quad .
\eeq
The extra $Z(2)$ variables ($z(l)$) that occur on the Wilson line
correlation function can be absorbed in the Haar invariant measure,
and the Wilson line correlation function reduces to the correlation
function of two fundamental Wilson lines, apart from some
normalization factors that arise because of the absolute value
of the $\lambda(p)$ variables. Since the extra $Z(2)$ variables can be absorbed
away, the action of the $SO(3)$ LGT reduces to that of the $SU(2)$ LGT.
In the $SU(2)$ LGT, at high temperatures,
the correlation function of two fundamental Wilson lines approaches
a non-zero value at large distances. This is precisely the behaviour seen for
the tiled Wilson line correlation function in the \Laa \ positive state.

So far the results were for asymmetric lattices. We now record some
observations on symmetric lattices. In the infinite lattice size limit,
a symmetric lattice corresponds to the bulk zero-temperature system.
However, simulations are always done on finite lattices.
A finite symmetric lattice can also be regarded as a finite
temperature system whose spatial volume is small (since $N_{\sigma}
\approx N_{\tau}$).
When the spatial volume
is small, the tunnelling probability between metastable states will
increase (since it goes as $\exp (-\alpha V)$ where $\alpha$ is some
positive constant and $V$ is the volume). The simulations on a symmetric lattice are
more likely to see tunnellings between metastable states and this
is indeed the case.
For large $\beta_{a}$ (3.5), the state with $L_{a}$ negative also
appears whenever the simulation is begun from a hot start. The state with
a cold start rarely settles down to a steady value and oscillates as shown
in Fig.~\ref{oscill}. 
This behaviour occurs for many couplings
(and seeds of the random number generator ) and is not a feature of
any particular run or updating algorithm. 
Also it occurs only for large symmetric 
lattices and is never observed on, for instance, an $N_{\tau}=3$ lattice.
It is natural to interpret this oscillation as
tunnelling between
degenerate or almost degenerate metastable states. 
If this is indeed the case, we can study the tunnelling probability between
these states. This suggests a small experiment.
So far, the temporal extent of the lattice was kept fixed at $N_{\tau}=3$
and the temperature was varied by varying $\beta_{a}$. We now fix $\beta_{a}$
and vary $N_{\tau}$. This has the effect of varying the temperature at a
fixed coupling (which can be chosen to be large). The reason for doing this
is as follows: fixing $\beta_{a}$ and varying $N_{\tau}$ not only has the
effect of varying temperature, but if $N_{\sigma}$ is kept fixed, it also 
has the effect of achieving a simulation in a small volume. This will
aid tunnellings between metastable states which are degenerate or
almost degenerate.
We choose two values of $\beta_{a}$; they are $3.5$ and $8.5$.
The purpose of this exercise is to see how the tunnelling 
probabilities are
affected as one increases the temperature.
We find it convenient to plot the distribution function of $L_{a}$
(now this refers to the value of the adjoint Wilson line averaged over
all lattice sites) as a function of $N_{\tau}$ ($\beta^{-1}$). This
evolution is shown in Fig.~\ref{vsnt1}. On lattices of large temporal
extent, one sees two peaks in the distribution of $L_{a}$ and these are
centered on positive and negative values.
One notices a gradual movement
of density from the $L_{a}$ negative region to the $L_{a}$ positive region
as $N_{\tau}$ is decreased. For $N_{\tau} =4$, the two peaks have
disappeared and there is only a single peak over the $L_{a}$ positive state.
This same experiment is repeated in Fig.~\ref{vsnt2} 
for a higher coupling ($8.5$) and one again observes a movement of
density from the $L_{a}$ negative  region to the $L_{a}$ positive
region, but this time the transition to a single peak occurs at a
larger value of $N_{\tau}$ ($N_{\tau}=5$).
This suggests that the $L_{a}$ positive state appears at higher values
of $N_{\tau}$ for larger values of $\beta_{a}$.
The above distributions were plotted after gathering 
data from 50000 iterations.
We have also studied the densities of
$L_{f}$ at a single site as the temperature is increased and
we again observe the shape changing from
a single peak centered on zero to a double peak symmetrically
distributed about zero \cite{thesis}.

Finally we wish to make a few remarks about the shift in the 
critical value of $\beta_{a}$ as a function of $N_{\tau}$. The bulk
transition moves to $\beta_{a} \approx 2.4$ on an $N_{\tau}=2$ lattice
and is at $\beta_{a}\approx 2.52$ on a $N_{\tau}=3$ lattice. We have
not observed any significant shift on an $N_{\tau}=4$ lattice.

This concludes our numerical studies of the $SO(3)$ LGT. 
Before we interpret our numerical results, we would like to point out
an important relation between the couplings of the $SU(2)$ and 
the $SO(3)$ LGT
at weak coupling.
 The relation between $\beta_{f}$ and $\beta_{a}$ when
both are large is
\beq
\beta_{f}=\frac{8}{3} \beta_{a} \quad .
\label{wcoup}
\eeq
This relation is true in the naive classical limit and does not
represent the effects of all the quantum corrections but it is still
a good guide to the weak coupling behaviour of the $SU(2)$ and the $SO(3)$ LGTs.
If there is a deconfinement transition in the $SO(3)$ LGT
at a large coupling, 
which is separated from the bulk transition, it must occur at $\beta_{a}>2.6$.
This means that the corresponding transition in the $SU(2)$ LGT must
occur at $\beta_{f}>5.6$ ( assuming that the weak coupling relation is
approximately valid at these couplings). Simulations 
in \cite{nt16} have shown that $\beta_{f}^{cr}=2.76$ on a $N_{\tau}=16$
lattice. This would mean that deconfinement transitions in the
$SO(3)$ LGT require very large temporal lattices.
Such large temporal lattices correspond to very low temperatures (in lattice
units).

We will now gather together all our numerical observations and 
try to tie them
up to arrive at a consistent
physical picture. We will find that this presents
several difficulties and that we are faced with many possibilities.
The discontinuous behaviour of the plaquette square and the
$Z(2)$ monopole density at $\beta_{a}\approx 2.5$ appears to be a
replica of
the transition in the bulk sysytem. It looks as if finite temperature
effects hardly shift the $Z(2)$ transition. As this 
transition is so
similar to the bulk transition, we expect that only the $Z(2)$ degrees
of freedom are changing across it. 
Hence we expect the
confining properties of the gauge theory to be unchanged across this
transition.
That the two phases differ by a distribution of the $Z(2)$
degrees of freedom is also clear from Fig.~\ref{sigdis}. These observations
are also in line with the studies made in \cite{hall}.

We now come to the behaviour of the adjoint Wilson line which is the
most striking aspect of our numerical results.
At low temperatures, the adjoint Wilson line has a very small
numerical value (infact it is very close to zero all the way till the
transition).
This small value at low temperatures is
quite unexpected because a static quark in the adjoint representation can always
bind with a gluon and form a state of finite energy.
Though we expect the adjoint Wilson line to be always non-zero, there is no
reason why it should take such a small value at low temperatures.
Nonetheless, in the
strong coupling approximation,
\beq
\langle L_{a}\rangle \approx (\beta_{a}/3)^{4 N_{\tau}} \quad,
\eeq
and is quite small on the lattices that we are using. This explains the small
value of \Laa \ in the strong coupling region though it does not provide a reason
why \Laa \  should be small all the way till the phase transition.
The adjoint Wilson line taking two distinct values at high temperatures
is the most unexpected feature of our results.
Equally puzzling, is the observation that observables
like the plaquette square appear to take 
almost the same values in these two states.
This behaviour is very reminiscent of the
high temperature phase of the $SU(2)$ LGT in which
one observes metastable states which are related
by a $Z(2)$ transformation.
In the $SU(2)$ LGT,
the global $Z(2)$ symmetry ensures that
the two states have the same free energy. 
In the $SO(3)$ LGT, there is no obvious symmetry relating the \Laa \
positive and \Laa \ negative states;
the presence
of two degenerate minima in the free energy in the absence of any 
such symmetry 
would be quite a remarkable instance. Moreover, although
the physical interpretation of the \Laa \ positive state is quite clear- it is
similar to the deconfined phase of the $SU(2)$ LGT,
the \Laa \ negative state does not easily admit a physical
interpretation.
However, as both these states seem to appear immediately after the
bulk transition, we are led to the different possibilities
considered by \cite{gav} in their studies of the
mixed action $SU(2)$ lattice gauge theory. These include,  (i) only bulk transition
and no deconfinement transition, (ii) only deconfinement transition and
no bulk transition, and (iii) two transitions with
a separation which we are unable
to resolve with our numerical methods. 
The possibility (i) goes against many theoretical and numerical
arguments favouring a deconfinement transition at high temperatures.
(ii) requires the transition to be of second order according 
to the arguments
in \cite{yaffe}. It also requires the transition point to shift
as the temporal lattice size is increased. We have not noticed any
significant shift in the critical coupling from $N_{\tau}=3\ to\ 4$.
Though we do not have a proof for any of these possibilities,
the third
possibility seems to be the least dramatic of the three.
Before proceeding further to interpret our numerical results,
we take the point of view that the presence
of the \Laa \ negative state is quite significant and has to be properly
accomodated in any scheme. Though the physical interpretation of
the \Laa \ positive state is quite clear, the \Laa \ negative state
still needs to be explained. 

Let us examine two possible interpretations.
One possible interpretation is that we have only the bulk
transition driven by the $Z(2)$ monopoles.
As we have observed only one transition, and
the states with \Laa \  positive
and \Laa \  negative appear immediately after this transition,
this may seem quite a promising explanation. 
One may also think that these two states are physically equivalent.
A recent measurement in \cite{saum} of the correlation length 
of the adjoint Wilson line
came up with the result that it was the same in the
two states. This seems to support the picture of two distinct
but physically equivalent states. Nevertheless,
this interpretation does have its problems. Our studies of the plaquette
square observable and the tiled Wilson line correlation function
show that these two observables behave differently in the
two states. Apart from the slightly different numerical values
of the plaquette square operator in these two states (which
measures the energy density in the bulk system), the tiled
Wilson line correlation function differs drastically in the \Laa \ positive 
state and
the \Laa \ negative state.
Morover, the exact physical equivalence of these states, in the absence of
a symmetry relating the two, would be quite a remarkable instance.
We have not been able to discover any symmetry that maps the \Laa \ 
positive state to the \Laa \  negative state, and 
which leaves the action invariant.
In the absence of such an explicit symmetry transformation,
we cannot be sure that
there does indeed exist such a transformation.
The phase diagram
of the $SO(3)$ LGT at non-zero temperature in this scenario, where there is
only a single bulk transition, 
would be as in
Fig.~\ref{pha} without the dotted line. There is only the bulk 
phase transition which is driven by the $Z(2)$ monopoles.

The other interpretation is that the states with \Laa \  positive and
\Laa \  negative are physically quite different. 
From a numerical standpoint, this seems to be 
supported by our observations of the plaquette square observable
and the tiled Wilson line correlation function, both of which 
behave differently in these two
states.
Also, a comparison of the distributions of
$L_{a}$ at high and low temperatures suggests that the \Laa \ negative
state and the \Laa $\approx 0$ (which is the low temperature confining phase) 
state are structurally similar, apart from the
width of their distribution functions (of $L_{a}$ and $L_{f}$).
Our observations on symmetric lattices also suggest
that this state can be associated with the bulk system.
A study of the tunnelling probabilities indicated that there was a
passage from the $L_{a}$ negative region to the $L_{a}$ positive
region as we increased the temperature.
This would suggest that the \Laa \ negative state is associated with the bulk
(confining) phase which passes into a deconfining phase at high
temperatures.
The tiled Wilson line correlation function also 
has the same behaviour
in the low temperature phase and the \Laa \  negative state.
An important consequence of this
interpretation is the
existence of a phase transition between the bulk
phase and the deconfined phase at large couplings,
which is quite different from the
$Z(2)$ transition.
Let us mention some of the questions raised by this picture.
If the two states are physically very different, why is it that they
always  seem to appear together? Within our present analysis we cannot
answer this question satisfactorily but it is possible that there 
are two minima in the effective potential 
at high temperatures which are closely spaced,
and this causes the configurations to get trapped in one of the 
two, almost
degenerate, minima. 
Also, we should emphasize that it is only the 
cold start that always ends up in the $L_{a}$
positive state; a hot start usually ends up in the $L_{a}$ negative
state.
A cold start (in which the initial $L_{a}$ is +3) 
is already close to the \Laa \  positive state and always
reaches that state. A hot start, on the other hand, corresponds to
$L_{f} \approx 0$ and usually settles to the \Laa \  negative state.
If the effective potential has two minima of different depths, then even a local
minima can appear as a very long lived metastable state. 
Such examples are known to occur in spin glass systems.
Since our
updating algorithm is a local algorithm, it will find it difficult to move
the system away from a local minimum. Morover, if there is no symmetry connecting
these two minima , it is difficult to make global updates
(like flipping of all the spins) which move the system from one minimum to the
other.
A mean field analysis \cite{sri} by one of the authors (S.C.) 
shows that the \Laa \  negative
state persists as a local minimum, even at high temperatures, and this may
explain its appearance in simulations even at large couplings.
Though we have not been able to directly detect a phase transition at
large coupling,
our observations of tunnelling probabilities
on lattices of large $N_{\tau}$ 
show that there is a passage from a double peak structure to a single peak
structure at high temperatures.
The argument after
Eq.~\ref{wcoup} also tells us that 
the search for this transition has to be carried out at very low
temperatures (large temporal lattices). 
In this picture, the phase diagram of the $SO(3)$ LGT would
be as in Fig.~\ref{pha}. 
The solid line is the $Z(2)$ driven transition which, 
at least on an $N_{\tau}=3$ lattice,
is a first order transition.
At zero temperature, both sides of the transition are confining phases and
at a non-zero temperature both phases undergo transitions to a common high
temperature phase.
The dotted line is the location
of the phase transition from the bulk phase to the deconfined phase.
This line lies very close to the $\beta_{a}$ axis as the
transition temperature is quite low. At large $\beta_{a}$, the line
will be similar to the line in the $SU(2)$ LGT as is expected from the
universality of lattice actions.
We consider this scenario to be more plausible
taking into consideration our numerical results and theoretical
expectations.

We now wish to discuss some theoretical issues which have an important
bearing on the interpretation of our results.
It can be shown that
the expectation value of 
the adjoint Wilson line is always a non-negative quantity. 
This basically follows from the fact that a static source in the adjoint
representation can always form a bound state with a gluon and give a positive
contribution to the partition function.
Infact, the free
energy interpretation of the average value of the adjoint Wilson line
in Eq.~\ref{freen}
presupposes that it is always a non-negative quantity. However, we seem to
be getting negative values in simulations. The
fundamental Wilson line in the $SU(2)$ LGT also takes positive and negative
values.
For the fundamental
Wilson line, one gets around this contradiction by saying that it is
the correlation function of two Wilson lines which can be given the
physical interpretation of measuring the free energy of a quark-antiquark
pair. This correlation function is always positive and there is no
problem with the free energy interpretation. 
The average value of an
isolated fundamental Wilson line on a finite lattice is in principle
always zero, because tunnelling between the two metastable
states always restores the
symmetry;
individual Wilson lines are
measured in simulations for
purely operational reasons.
The same avenue is not open for the adjoint Wilson line. Even an isolated
adjoint Wilson line can be non-zero, and it is always a non-negative
quantity. This seems to contradict the observations made in simulations
in which we have observed negative values for the adjoint Wilson line.
The way out is that in finite systems, there will always be tunnellings
(though one may have to wait a very long time) between the metastable
states  (which in this case are not connected by any symmetry).
In the $SO(3)$ LGT, the tunnellings are between the $L_{a}$
positive and the $L_{a}$ negative states, and since these states are non-uniformly
distributed about zero, they can give a net positive 
value for the average of $L_{a}$. 
This
is clearly seen in Fig.~\ref{vsnt1} and Fig.~\ref{vsnt2} where the
distributions of $L_{a}$ ensure that the mean value of $L_{a}$ 
is always in the positive region.
The reason why we have called the states observed in numerical
simulations as metastable states is that their thermodynamic significance
is not obvious.
Though we observe states having a positive and
a negative value of $L_{a}$ in numerical simulations, the average value
of $L_{a}$ is got by averaging over these two metastable states.
In the
thermodynamic limit, the
value of $\langle L_{a} \rangle$ in any phase is always a non-negative quantity.
If 
there is a phase transition 
at high temperatures, the observable that detects the transition is
$L_{a}$. Fig.~\ref{vsnt1} shows that average of $L_{a}$ need not change
discontinuously even though there is a multiple peak structure for $L_{a}$ 
across the phase transition. This is because the mean value of $L_{a}$ (which
is always a positive quantity) gradually increases as the temperature is
increased.
We now make a few remarks about the continuum limit.
Apart from the thermodynamic limit, one also has to take the continuum
limit so that the lattice system goes over to some physical system (in this case,
the Yang-Mills theory).
This
requires
taking the simultaneous limits, $N_{\tau}\rightarrow \infty$ and $a\rightarrow 0$,
and the passage to this limit can also affect the physical properties of the 
lattice system; the order
of the phase transition can also change as we approach the continuum limit
and may even become second order.
The possibility of a second order phase transition in the continuum
limit is also indicated by the fact that the absolute value of the adjoint
Wilson line decreases as the temporal lattice size is increased. Thus, the
multiple peaks seen in the adjoint Wilson line will move closer to each other
on very large temporal lattices. We conjecture that
in this limit, these distributions will resemble the corresponding
distributions in the $SU(2)$ LGT.

In this paper we have mainly emphasized the numerical results of our studies
of the $SO(3)$ LGT. We have observed a deconfining phase at high
temperatures which is just like the deconfining phase of the $SU(2)$
LGT. As there is no global $Z(2)$ symmetry operating in this model,
this is a deconfining phase without any symmetry breaking as in the $SU(2)$
LGT. We have also observed the bulk transition which is driven by
the $Z(2)$ degrees of freedom. In the course of our studies, we have
stumbled into a new metastable state which would have been apriori
very difficult to guess. The incorporation of this new state in the
model presents us with several difficulties; and a reconciliation 
of the numerical
observations with our physical intuition
leads us to consider different scenarios.
We have pointed out two possible scenarios for the phase diagram of
the $SO(3)$ LGT. Both of them are able to explain some of the
observations made in simulations, but they also pose problems for
a complete reconciliation between numerical observations and
physical expectations. Our analysis does show, however, that the $SO(3)$ LGT
has a much richer behaviour than the $SU(2)$ LGT.
It is quite likely that these features persist in
systems which have bulk transitions, and which are also
expected to have finite temperature deconfinement transitions.

One of the
authors (S.C.) has tried to make some analytical calculations in order
to explain the puzzling features of the $SO(3)$ LGT \cite{sri}. A
mean field analysis of the $SO(3)$ LGT
reveals the presence of the \Laa \ negative and \Laa \ positive 
states at high temperatures. The
structure of the metastable states in simulations, at high and low temperatures, 
can also be
explained by the mean field theory. The mean field theory
analysis also predicts the existence of a phase transition at large
$\beta_{a}$ for the $SO(3)$ LGT. 

One of the authors (S.C) would like to acknowledge useful discussions with
Rajiv Gavai and Saumen Datta.

\end{section}

\newpage
\begin{figure}
\label{trans}
\caption{ 
(a) plaquette square ($P$) and (b) $Z(2)$ monopole density ($\rho$), 
as a function 
of $\beta_{a}$. There is an abrupt rise in $P$ and a fall in $\rho$,
at $\beta_{a} \approx 2.5$.}
\end{figure}
\begin{figure}
\label{la}
\caption{(a)The two metastable states for $L_{a}$ at $\beta_{a}=3.5$.
The positive value 
of $L_{a}$ is reached
after a cold start, and the negative value is reached after a hot start.
(b) The variation of \Laa \  with $\beta_{a}$.}
\end{figure}
\begin{figure}
\caption{The distribution of $L_{f}$ and $L_{a}$ in the
\Laa \  positive state.\  $\beta_{a}=3.5$.}
\label{plus}
\end{figure}
\begin{figure}
\caption{The distribution of $L_{f}$ and $L_{a}$ in the 
\Laa \ negative state.\  $\beta_{a}=3.5$.}
\label{minus}
\end{figure}
\begin{figure}
\caption{The distribution of $L_{f}$ and $L_{a}$ in the low temperature phase
.\  $\beta_{a}=2.0$.}
\label{zero}
\end{figure}
\begin{figure}
\caption{The distribution of $L_{f}$ and $L_{a}$ in the high temperature phase of the
$SU(2)$ theory.\  $\beta_{f}=4.5$.}
\label{su2high}
\end{figure}
\begin{figure}
\label{sigdis}
\caption{ The distribution of $\lambda(p)$ (called $Z$ in the figure) 
 in the, (a) low and (b) high
temperature phases.}
\end{figure}
\begin{figure}
\label{evolu}
\caption{Plaquette square evolving with Monte-Carlo sweeps. The values of
$\beta_{a}$ are, (a) 8.5 and (b) 10.5.}
\end{figure}
\begin{figure}
\label{tiled}
\caption{The tiled Wilson line correlation function
 in the, (a) $L_{a}$ negative phase,
(b) $L_{a}$ positive phase, and the (c) low temperature phase.\  $\beta_{a}=3.5$
in (a) and (b) and $\beta_{a}=2.0$ in (c).
}
\end{figure}
\begin{figure}
\caption{\Laa \ on a $7^4$ lattice as a function of Monte-Carlo sweeps for (a) hot start
and (b) cold start. \  $\beta_{a}=3.5$ }
\label{oscill}
\end{figure}
\begin{figure}
\caption{The distribution of $L_{a}$ as a function of $N_{\tau}$ at
$\beta_{a}=3.5$.}
\label{vsnt1}
\end{figure}
\begin{figure}
\label{vsnt2}
\caption{The distribution of $L_{a}$ as a function of $N_{\tau}$ at
$\beta_{a}=8.5$.}
\end{figure}
\begin{figure}
\caption{ Possible phase diagram of the $SO(3)$ LGT at finite temperature.
The solid line is the bulk transition. The dotted line is the deconfinement
transition.}
\label{pha}
\end{figure}
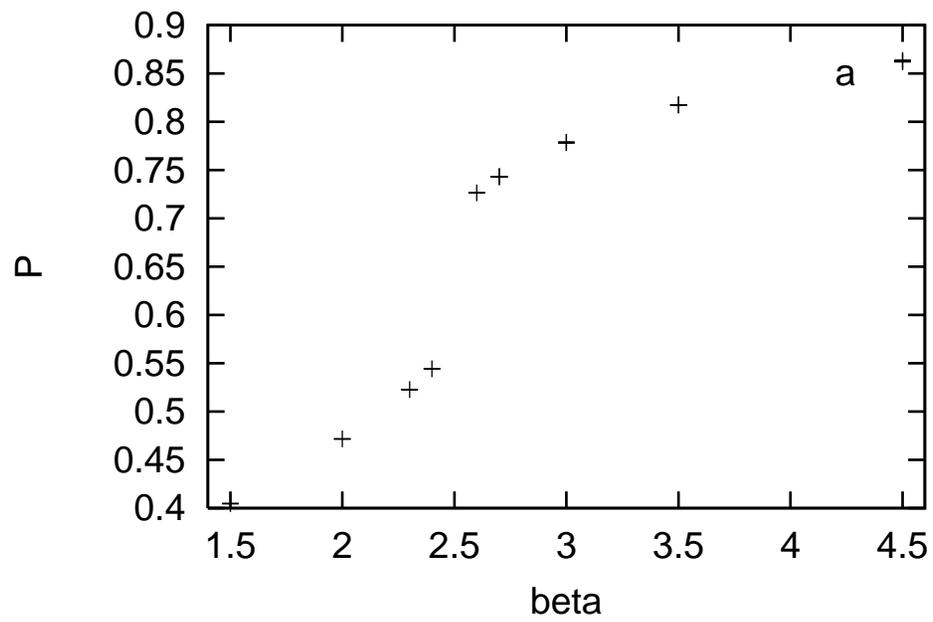
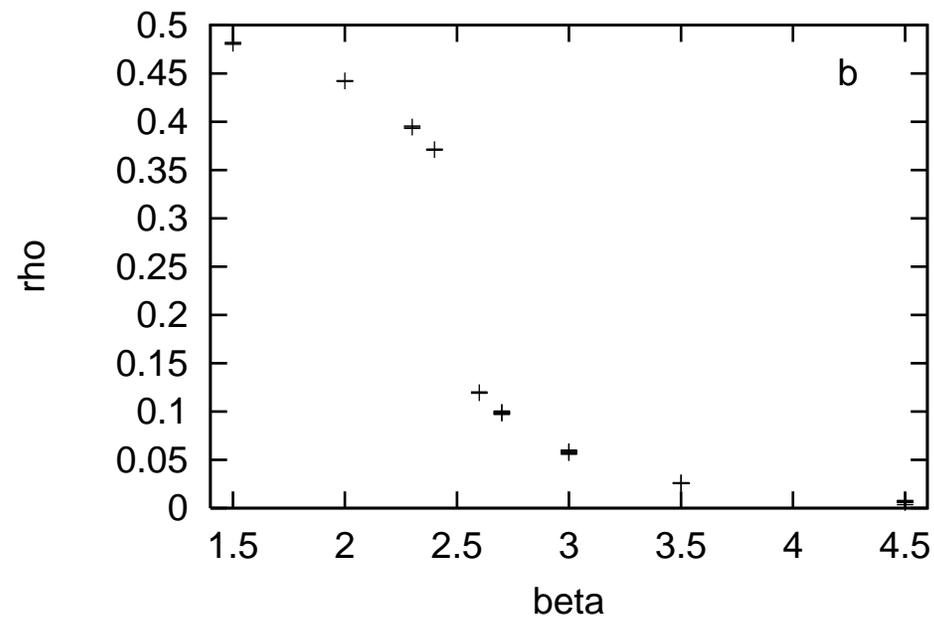
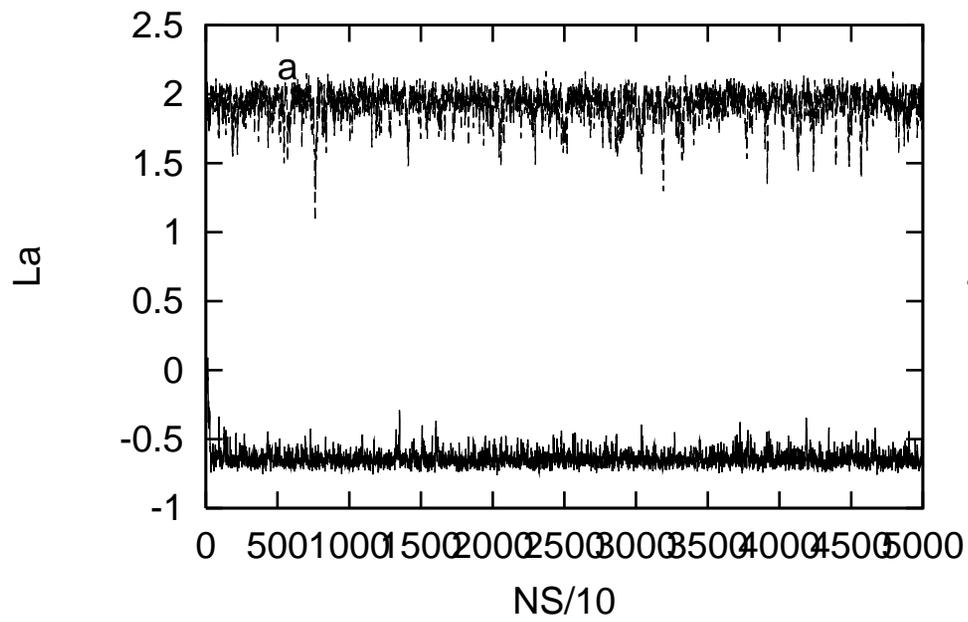
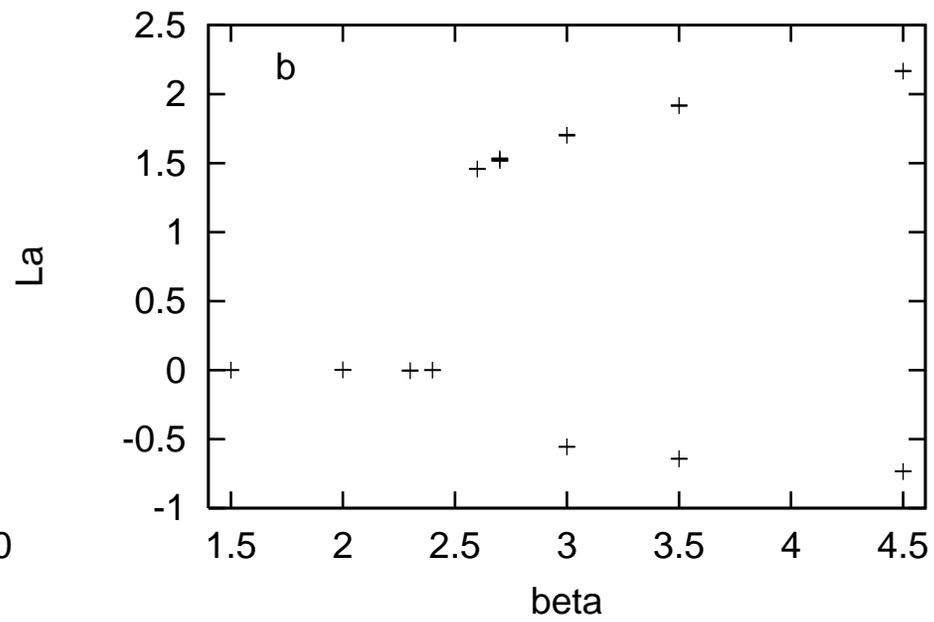
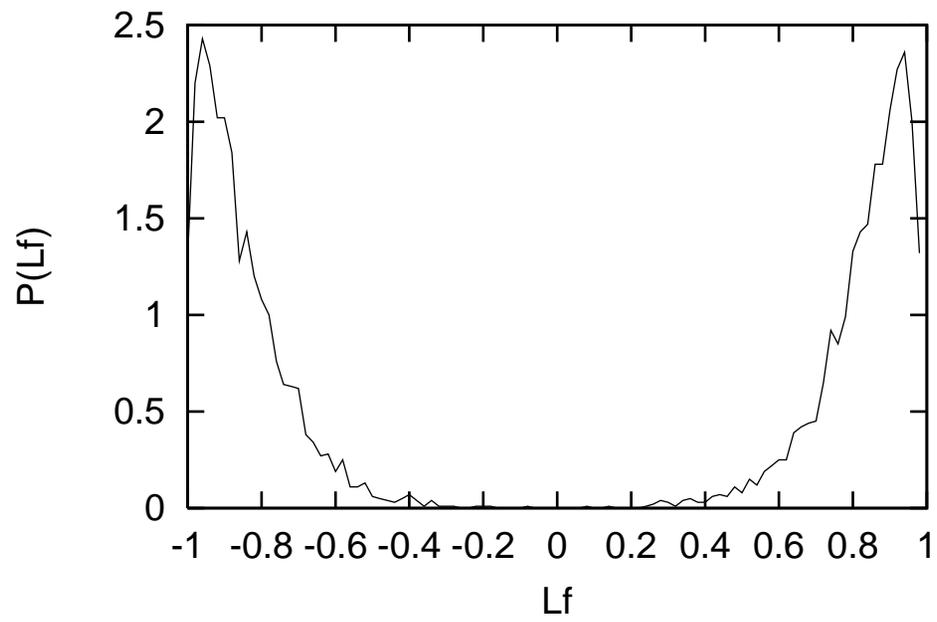
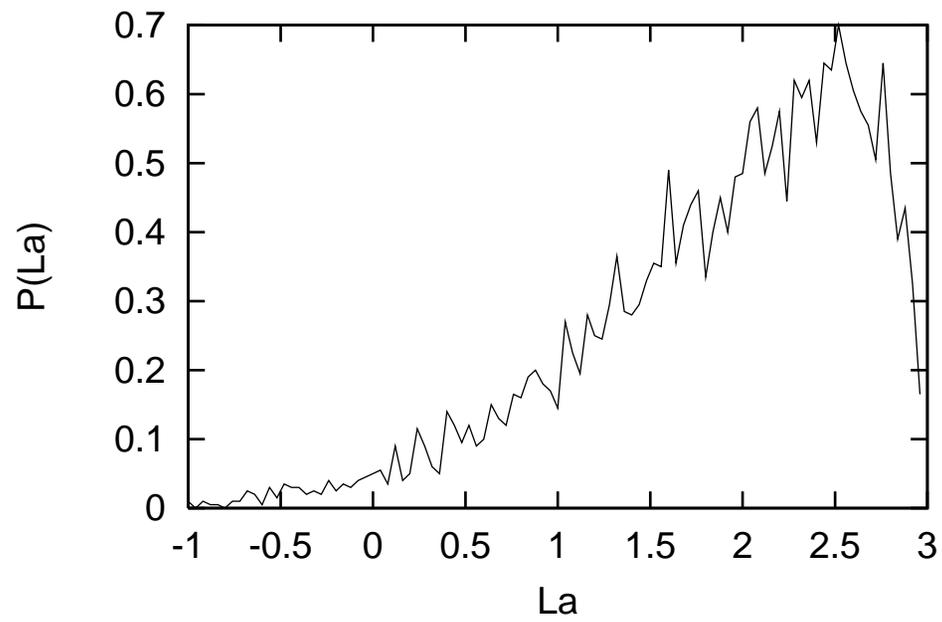
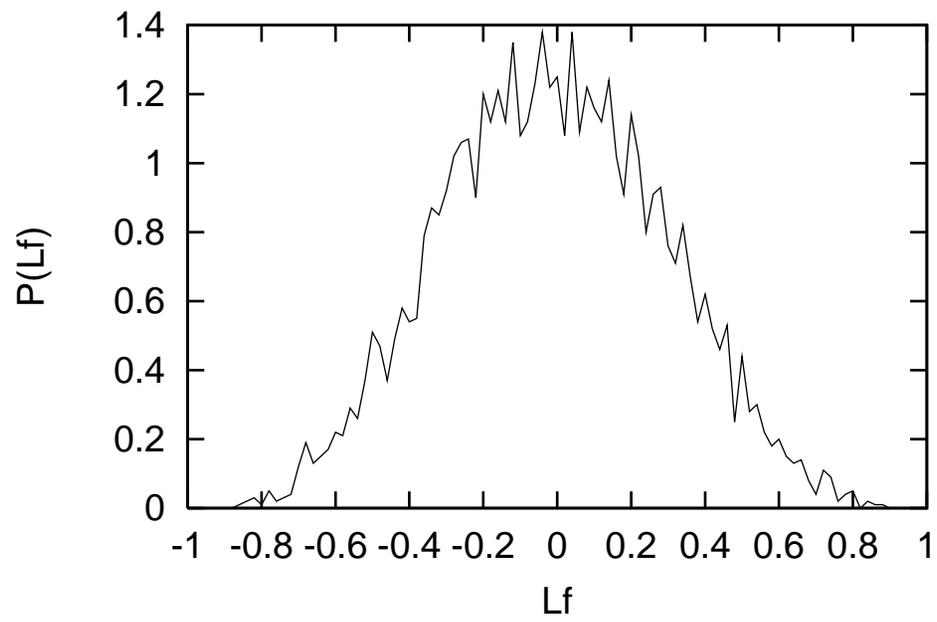
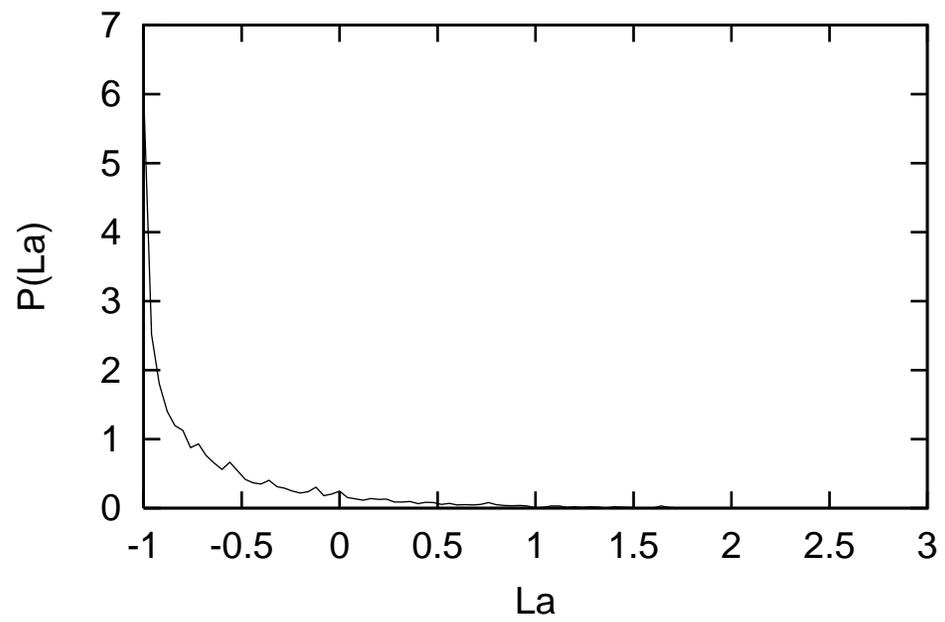
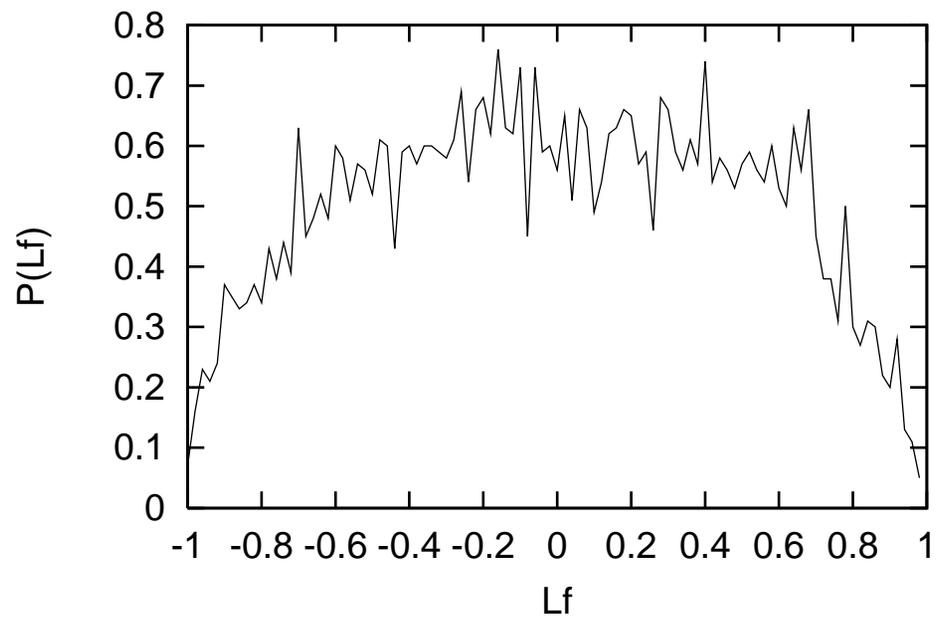
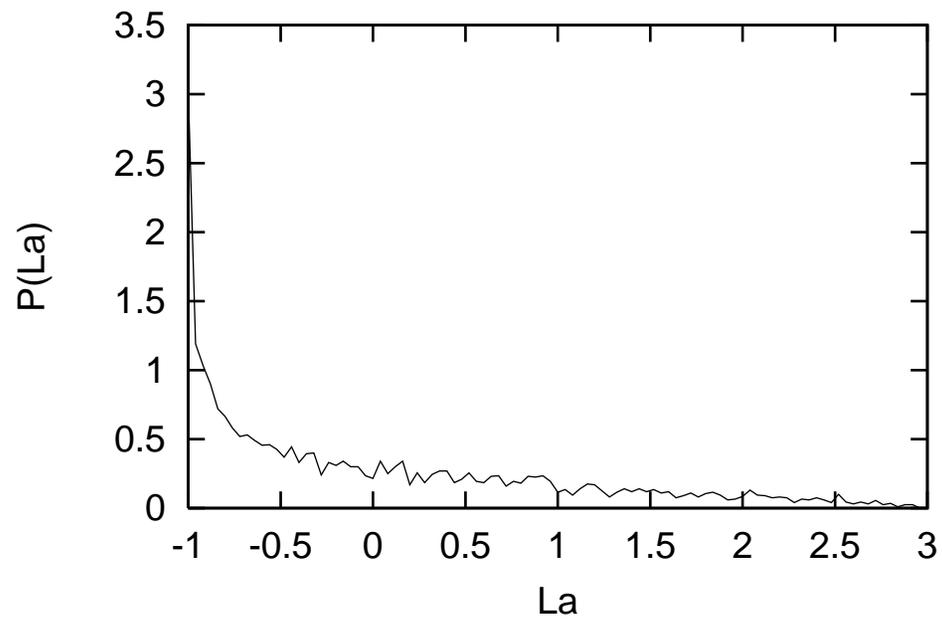
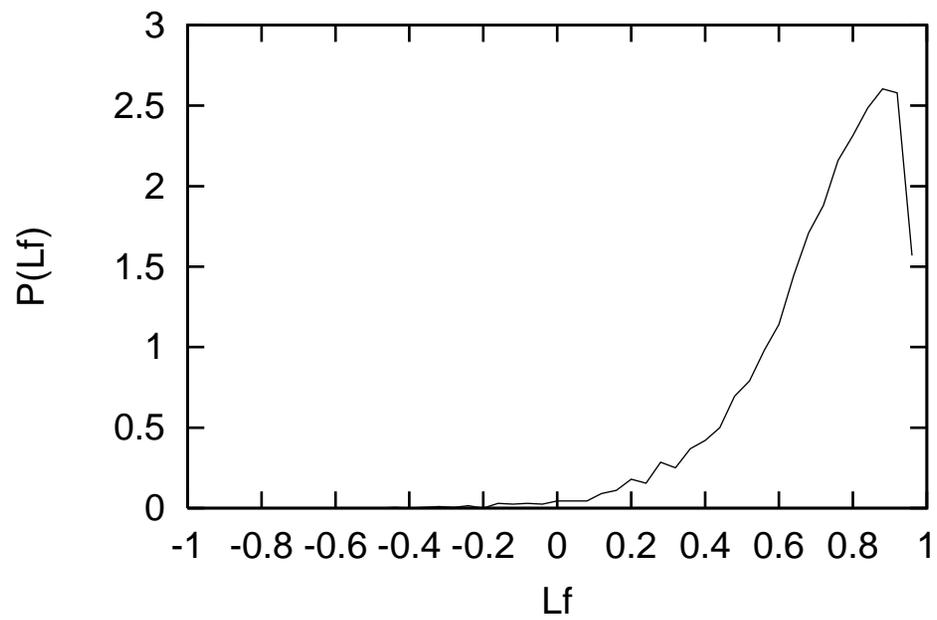
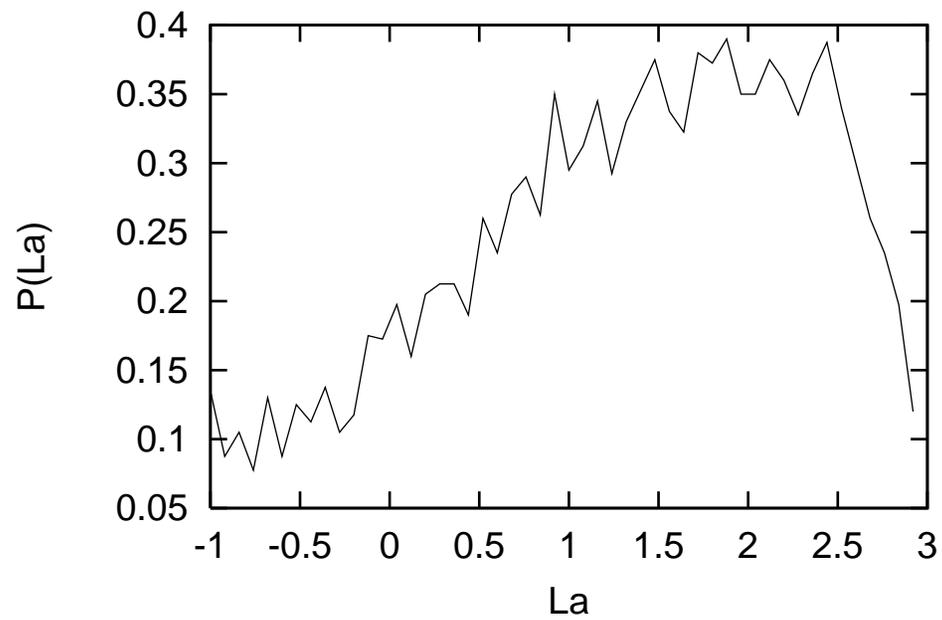
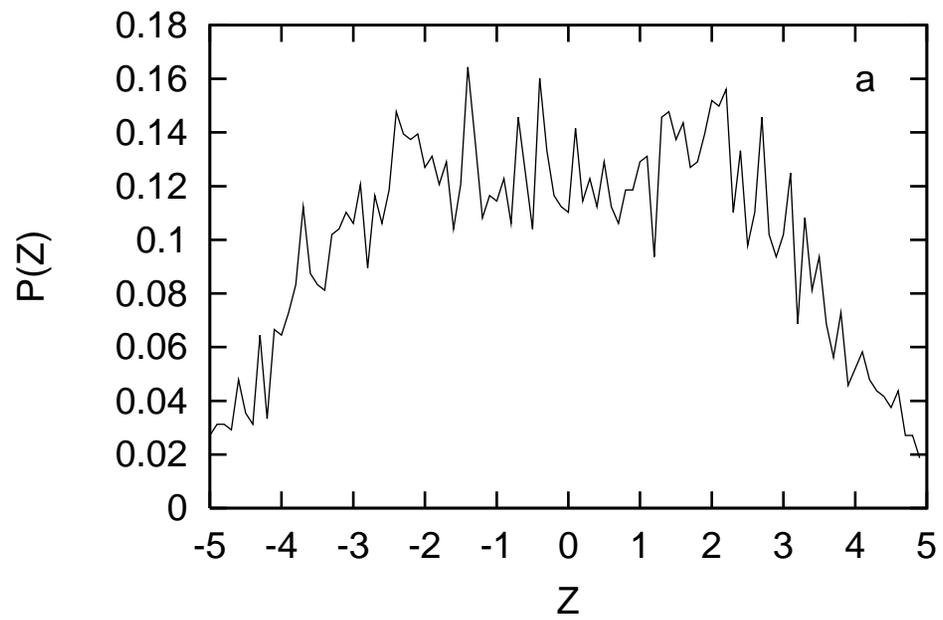
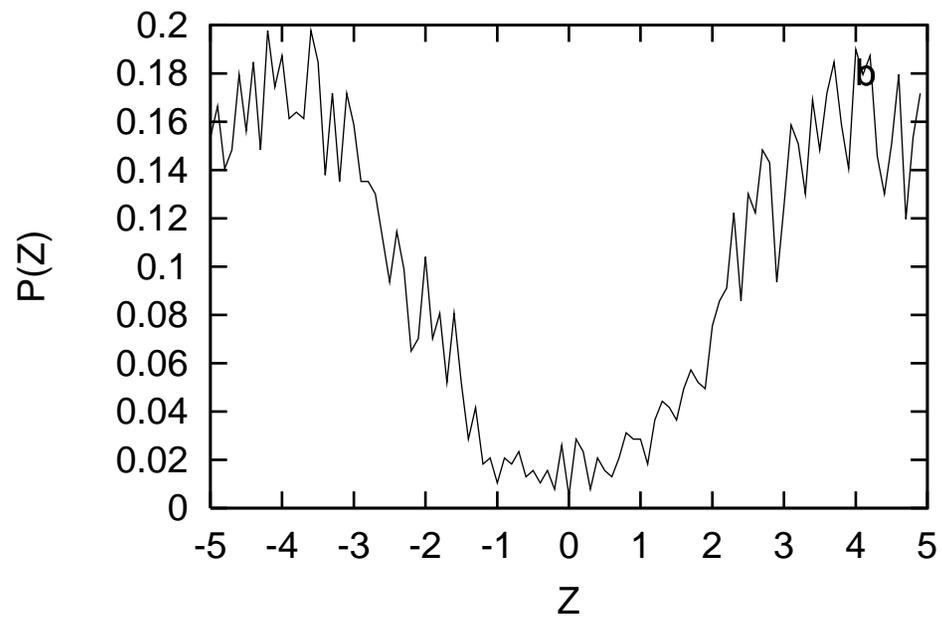
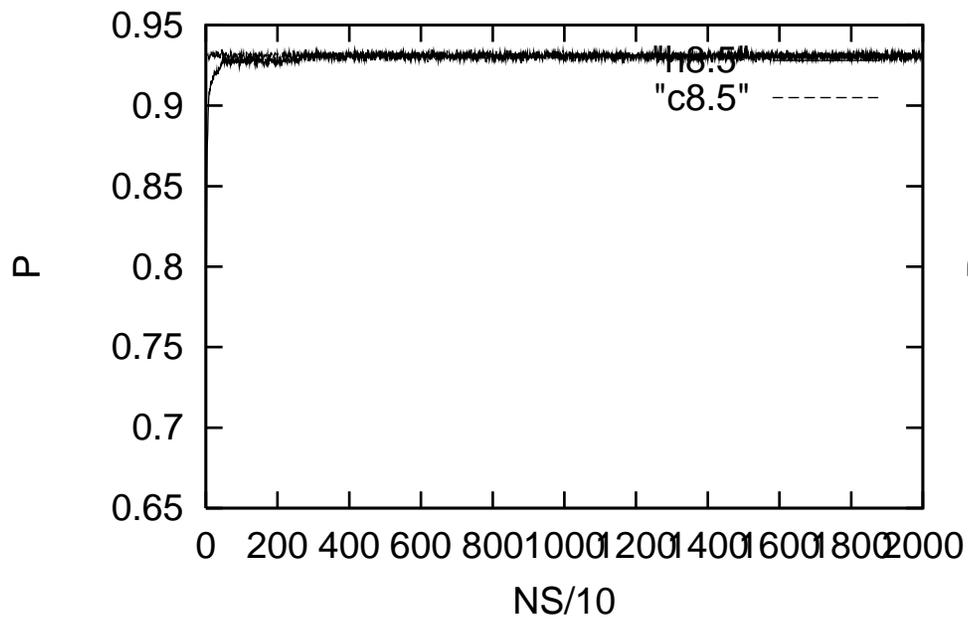
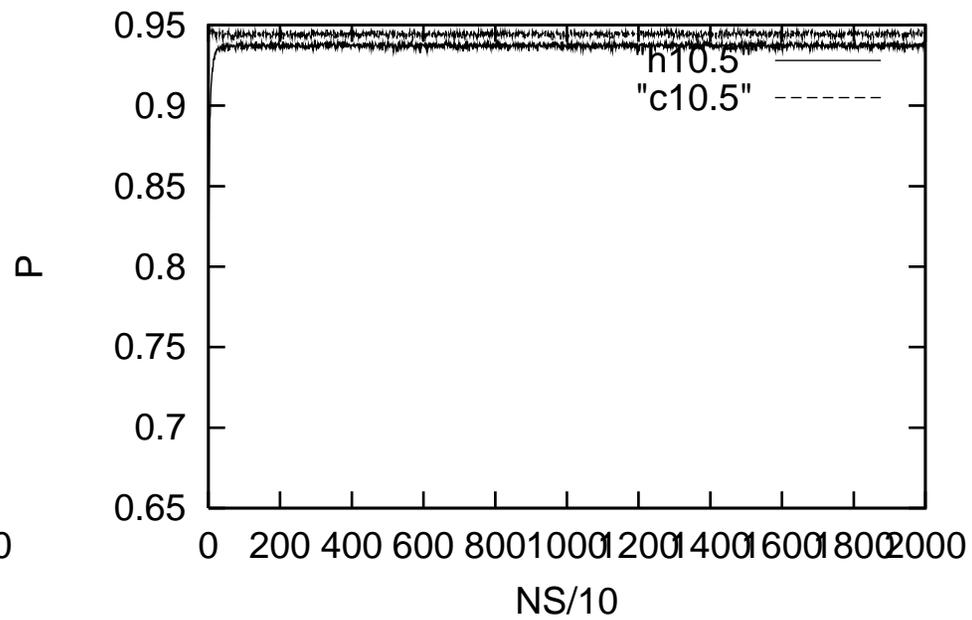
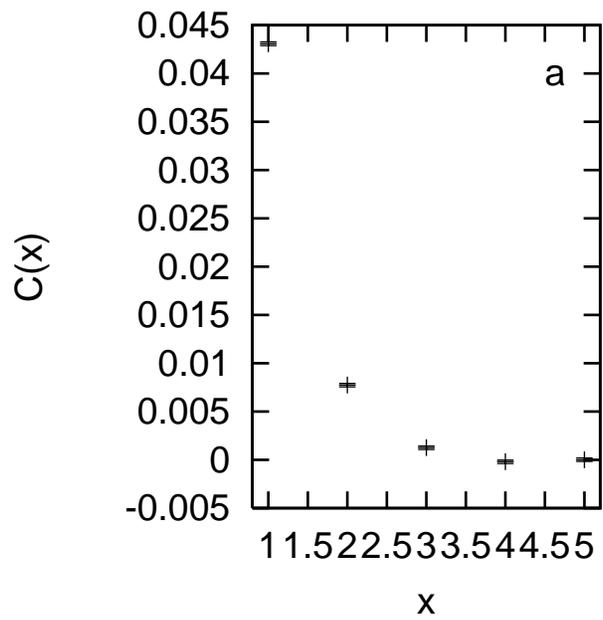
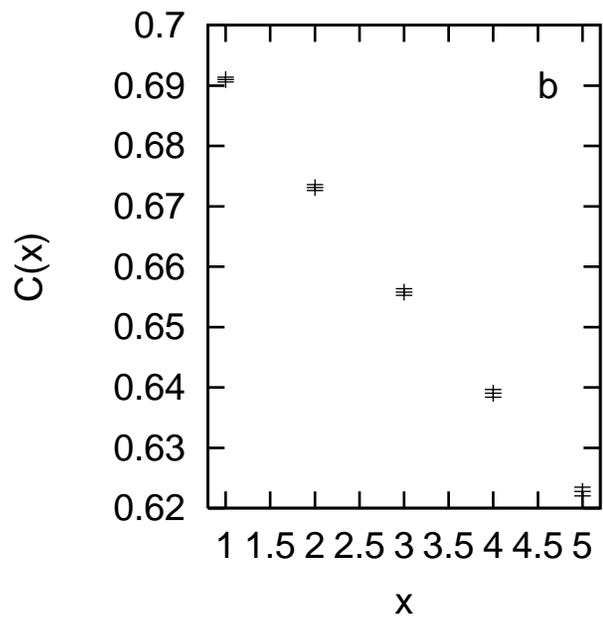
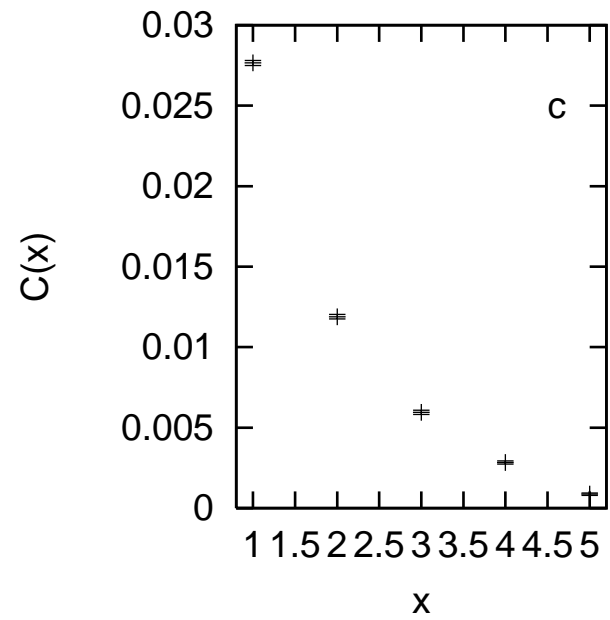
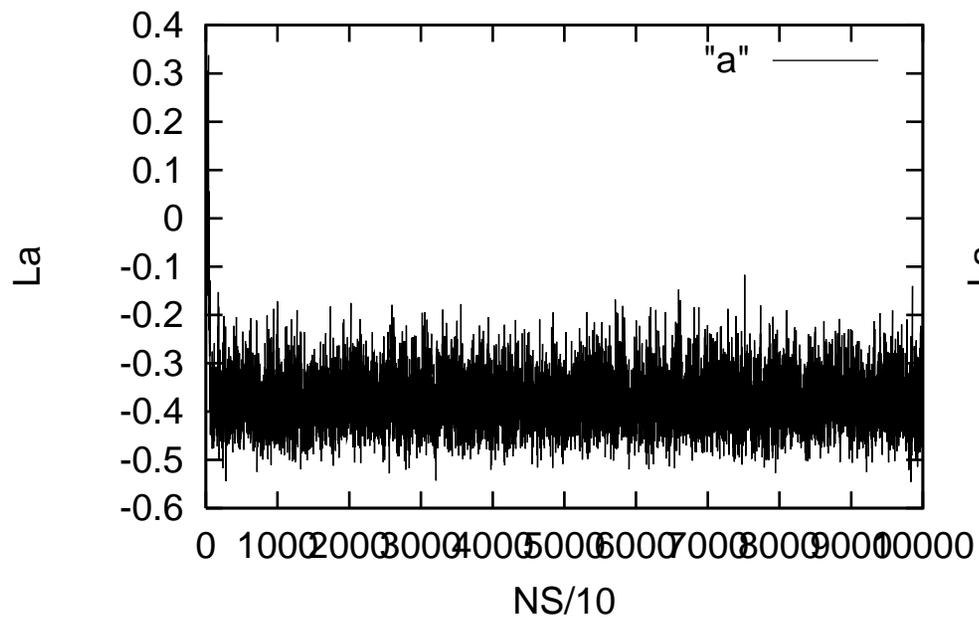
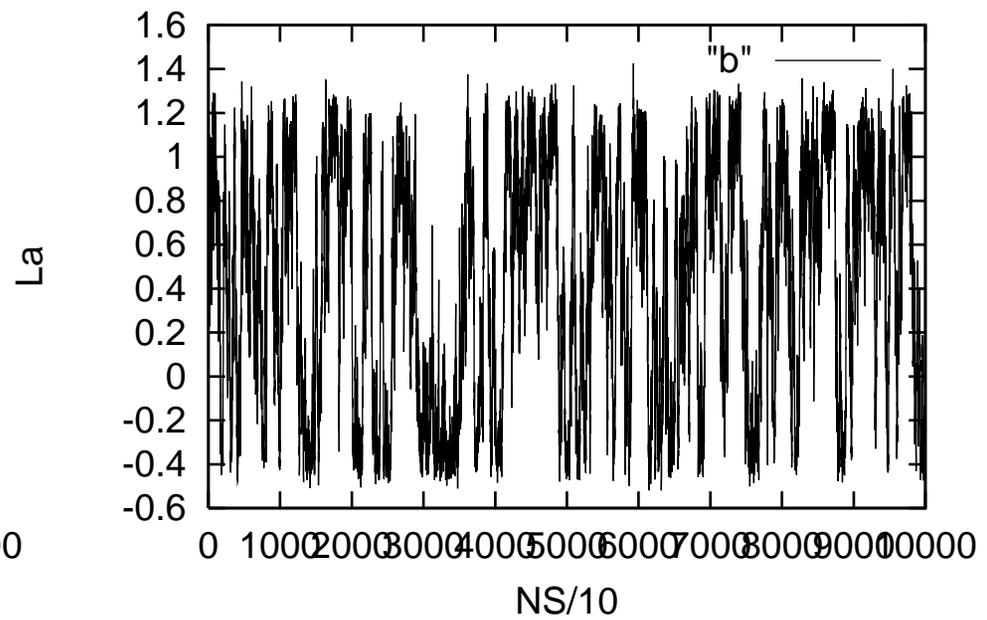
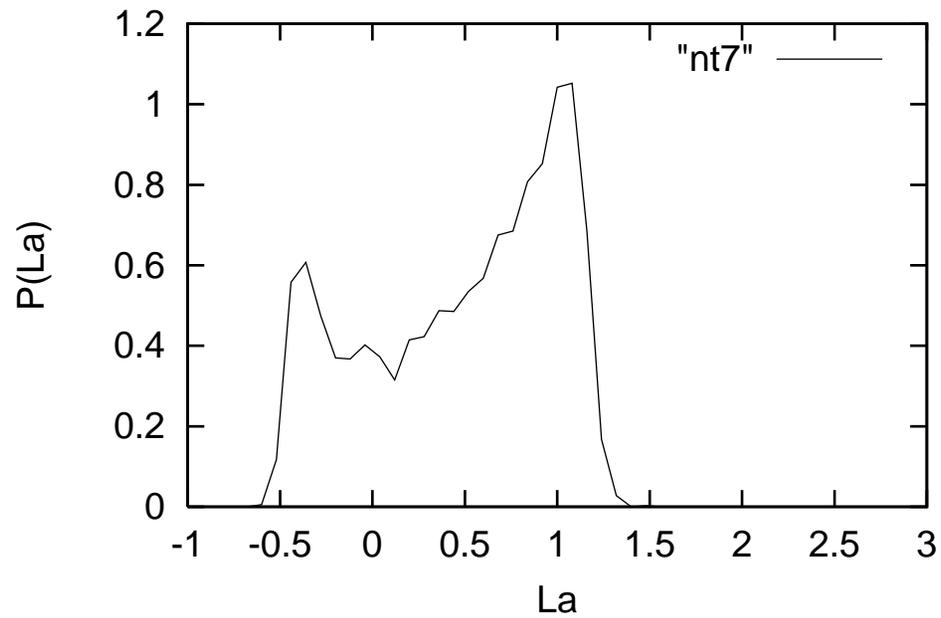
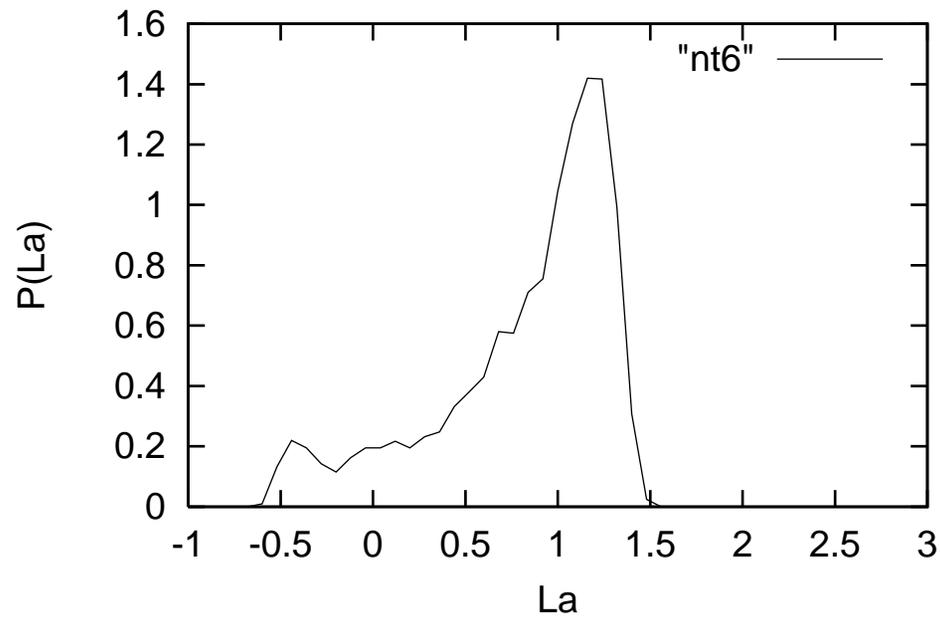
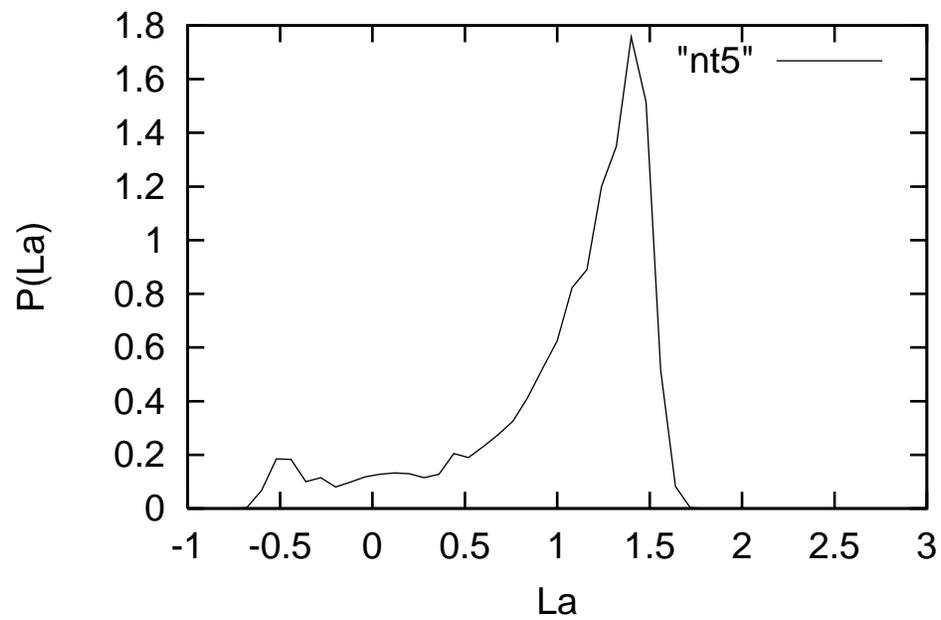
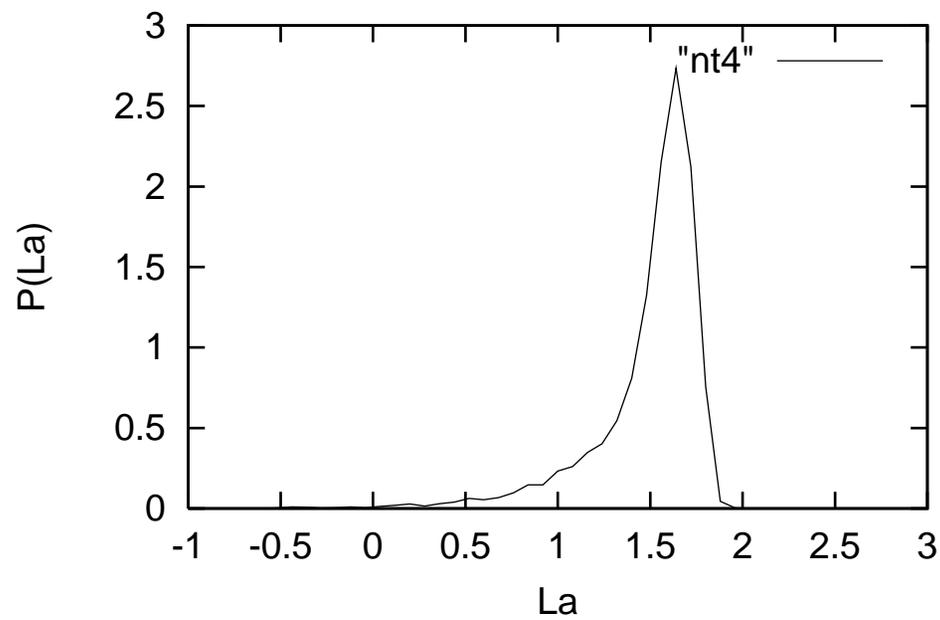
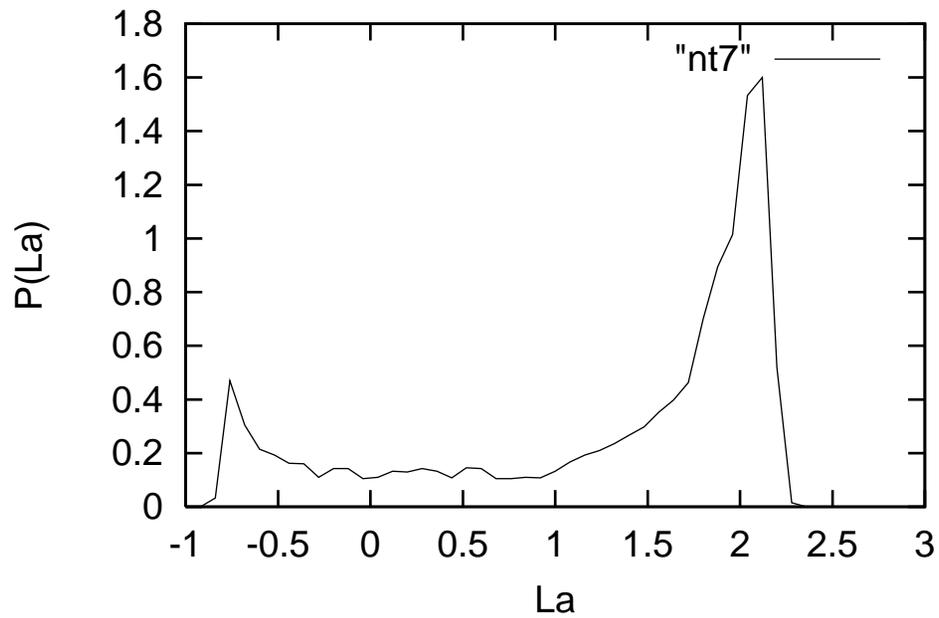
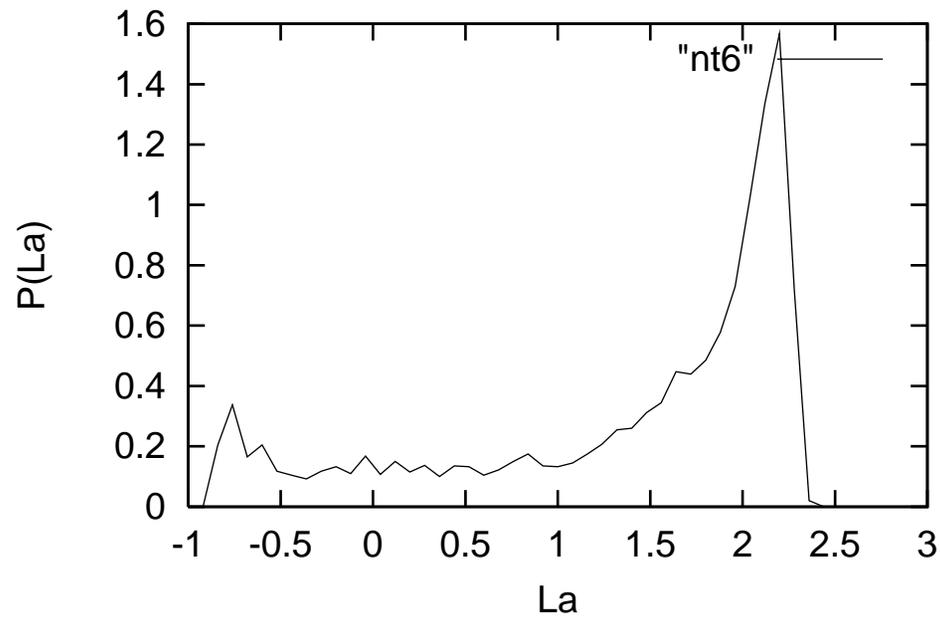
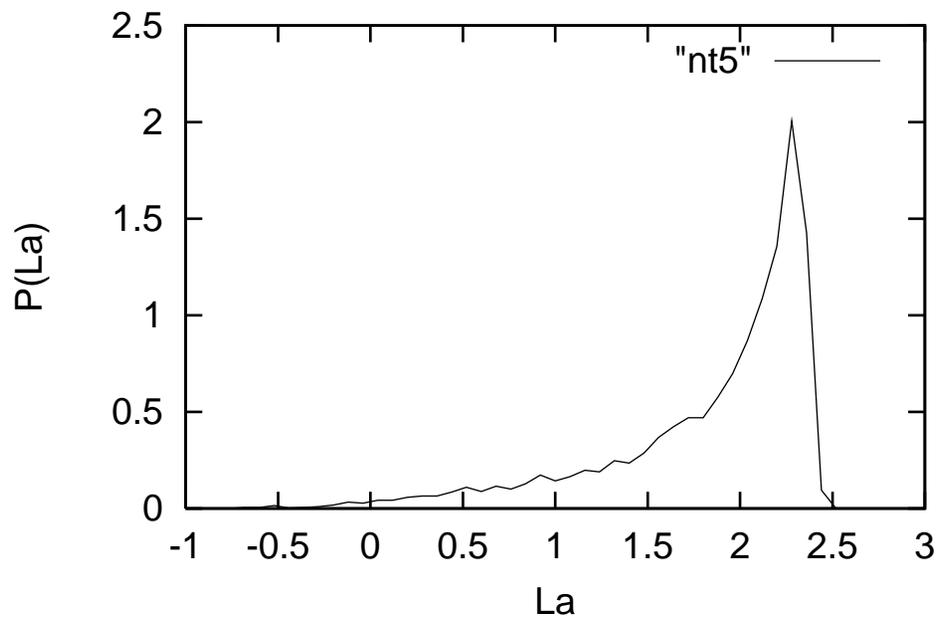
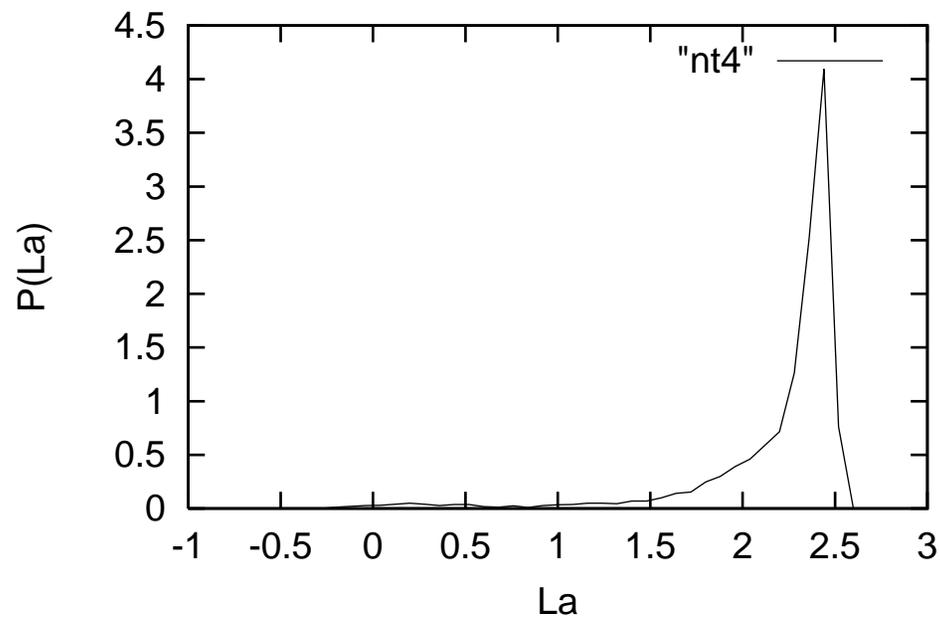
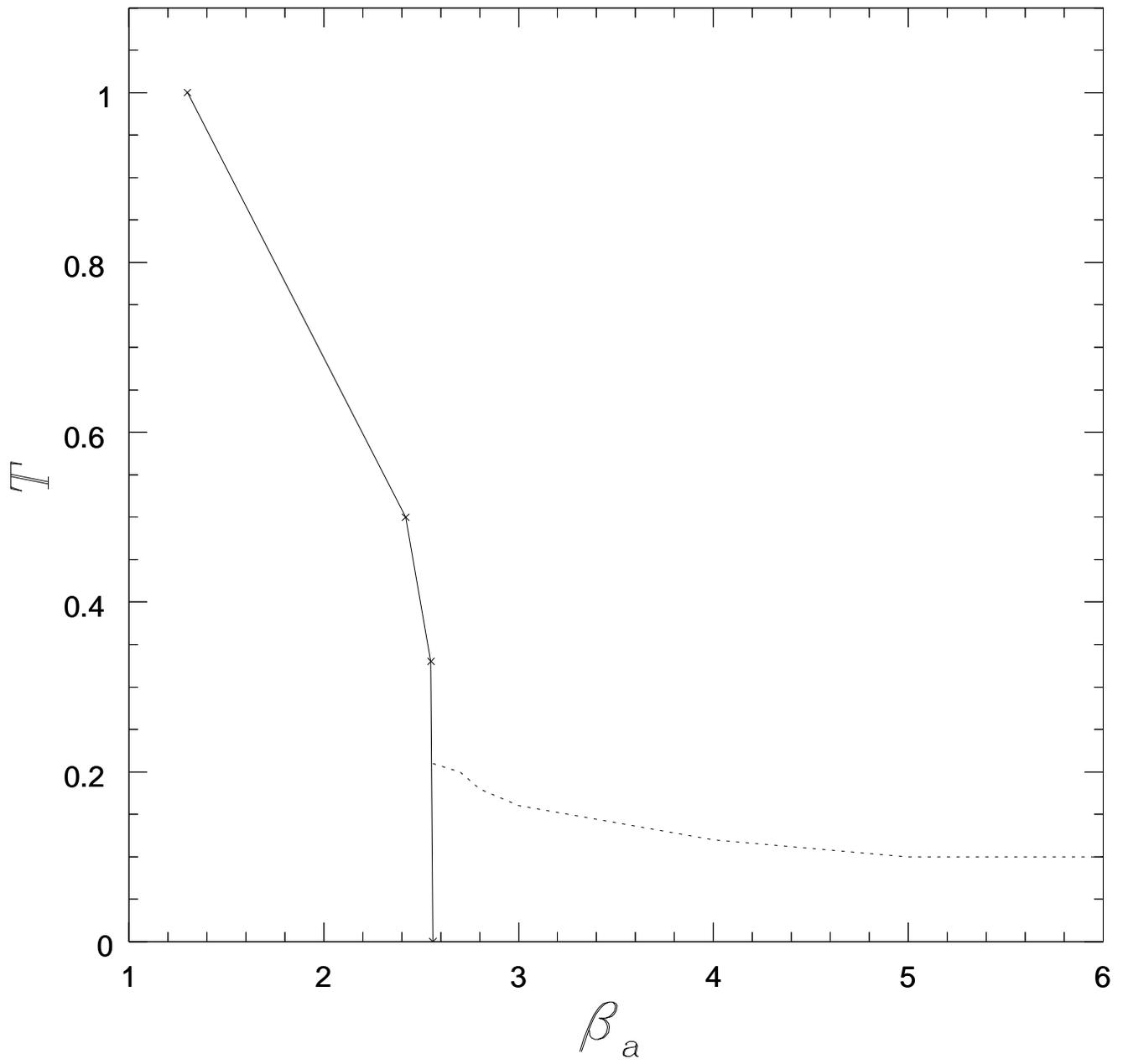
\end{document}